\documentclass[]{iopart}

\usepackage{iopams}  
\usepackage{epsfig}
\begin{document}

\title[]{In-Medium Baryon Interactions and Hypernuclear Structure}

\author{C.~Keil$^\dag$ 
\footnote{Christoph.m.Keil@theo.physik.uni-giessen.de}
, H.~Lenske$^\dag$ and C.~Greiner$^\dag$ 
}

\address{$^\dag$ Institut f\"ur Theoretische Physik I, 
               Heinrich-Buff-Ring 16, D-35392 Gie\ss en}

\begin{abstract}
We introduce a microscopic and relativistic theory describing free scattering and
finite nuclei for the octet baryons in a consistent quantum field theoretical
framework based on Dirac-Brueckner theory and nuclear mean-field. In a first 
quantitative approach - yet still heuristic for dealing with the $\Lambda$
hyperon - the quality
of the description of finite (hyper)nuclei based on free meson exchange potentials
is competitive with those from purely
phenomenological relativistic mean-field calculations.
In contrast to the latter our approach has the advantage of having a 
microscopic link to free interactions and
complete control on which classes of diagrams are included.
As a complementary way for determining hyperon-hyperon and 
hyperon-nucleon interactions the use of Hanbury-Brown-Twiss interferometry is 
discussed.
\end{abstract}




\section{Introduction}
For the study of baryonic systems at densities exceeding the nuclear matter 
saturation density $\rho_0$ by several times the understanding not only 
of the nucleon-nucleon 
interaction at high densities but also of the hyperon-nucleon and
hyperon-hyperon interactions is necessary.
Unlike the nucleonic sector 
-- which is fairly
well under control with high precision meson exchange potentials and 
Dirac-Brueckner theory --  
almost no scattering
experiments are feasible in the strangeness sector
 for principle reasons. An extension of baryon-baryon
 interactions to the S$\ne$0 sector on theoretical grounds is thus
necessary. A natural choice here is the use of SU(3)$_f$. 
Although SU(3)-symmetry is broken on the
level of the hadron masses, the relevant SU(3) charges strangeness and isospin are 
conserved by the strong baryon-baryon (BB) interaction. One can therefore expect that
on the level of the coupling constants this symmetry might hold to a 
reasonable degree
of accuracy, too. Nevertheless, even assuming exact SU(3) invariance for the BB
interactions does not yield an unique fixing of the relations between couplings
(as is already the case for the isospin in the NN sector). Points left
open by symmetry considerations are the pseudo-scalar/pseudo-vector ratio, the
$f/(f+d)$ ratio and the scalar sector. Especially the scalar sector requires a
separate effort of modelling since none of the particles of a scalar octet --
which can be constructed in accordance to the mesonic octets of pseudo-scalars 
and vectors -- is found in a way suitable to use it in the standard meson 
exchange approaches as a particle with sharp mass $m_s$. The analysis of NN
scattering data within the one-boson-exchange (OBE) framework shows however that
the inclusion of scalar exchange channels is required to describe the data
accurately. Such an analysis is, as mentioned earlier, not possible for the
S$\ne$0 sector, making the scalar channels subject to modelling. One way to 
generate a controlled extension is to introduce the scalar mesons 
dynamically as correlated
exchanges of pseudoscalar mesons in the $\pi\pi$ and $K\overline{K}$
channel. Such an approach is used e.g. by the J\"ulich potential 
\cite{Reuber:1996vc}.

The comparison to experimental data of such a SU(3) extended theory is needed to
put it on safe grounds. ``Simple'' hadronic systems containing strangeness that
can be produced and observed very accurately by experiment are hypernuclei. To
pin down or verify the couplings and the degrees of freedom in the SU(3) model
a microscopic translation of the free interactions to finite densities has to
be performed. This can be done by the use of Dirac-Brueckner theory incorporating
the effect of Pauli-exclusion and the dressing of the free baryon propagators
due to self-energies in the medium. In this approach the two particle in-medium
scattering equation is solved self-consistently in ladder approximation.
An application to finite systems is not possible for technical reasons. As an
approximation the DB self-energies can be used in the local density approximation
for finite nuclei. The density dependent relativistic hadron field theory, 
presented in the next section, provides such a framework 
\cite{Fuchs:1995as,Hofmann:2001vz,Keil:2000hk}.

\section{DDRH Theory}
\begin{figure}
\centering\epsfig{file=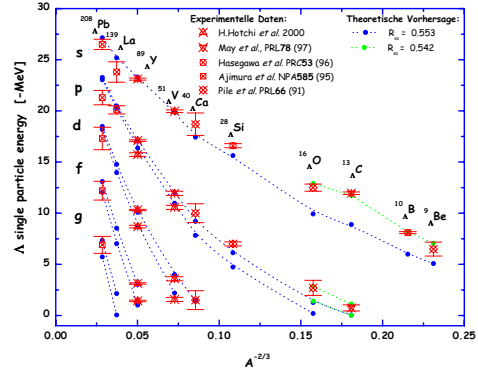,width=.5\linewidth}
\caption{Single particle spectra of single $\Lambda$ hypernuclei. Comparison
of DDRH calculations with experimental world data. \cite{Keil:2000hk}}
\label{fig:hnspec}
\end{figure}
The density dependent relativistic hadron field theory (DDRH) 
\cite{Fuchs:1995as} is a relativistic Lagrangian field theory with baryons
and mesons as degrees of freedom. It utilizes a mapping
of DB self-energies onto DDRH mean-field self-energies with density dependent
couplings
\begin{equation}
\left(\frac{\Gamma_i(\hat\rho)}{m_i}\right) \rho_i \equiv \Sigma_{DB}^i
\end{equation}
for finite nuclei with realistic interactions describing 
free scattering.
To retain a Lorentz invariant field theory the density dependent
couplings are constructed as functionals of Lorentz scalar combinations of the
baryon field operators $\hat\rho$. The description of isospin nuclei within 
DDRH has been
shown to be very good using the Bonn A and the Groningen NN-potentials
\cite{Fuchs:1995as,Hofmann:2001vz}.

Due to the functional structure of the vertices, containing the baryon field
operators, the equations of motion for the baryons do get additional 
and particular terms, the
rearrangement self-energies. They account for static polarization effects of
the particles in the medium and are given by
\begin{equation}
\Sigma_{rearr}\propto
\frac{\partial\Gamma_\phi(\hat\rho)}{\partial\hat\rho}\rho_\phi\phi
\end{equation}
where $\phi$ is
the meson field and $\rho_\phi$ the respective density appearing as source
term in the meson field equation. These terms ensure
the thermodynamical consistency of the theory.

The extension to the strangeness regime would be straightforward if there were
DB calculations available for the complete baryon octet. An analysis of the
DB equations shows, however, that the coupling functionals of the hyperons can
be approximated by the coupling functionals of the nucleons scaled by a
constant factor and taking the respective hyperonic density as the argument
\cite{Keil:2000hk}:
\begin{equation}
\Gamma_{Y\alpha}(\rho_Y) \approx R_{Y\alpha}\Gamma_{N\alpha}(\rho_Y),
\end{equation}
where $R_{Y\alpha} = \frac{\Sigma_{Y\alpha}(\rho)}{\Sigma_{N\alpha}(\rho)}
\approx \frac{g_{Y\alpha}}{g_{N\alpha}}$ and $g_i$ are the free couplings. 

We applied this prescription to the mean-field calculation of single $\Lambda$
hypernuclei \cite{Keil:2000hk}, 
taking the scalar scaling factor $R_{\Lambda\sigma}$ from calculations
of $\pi\pi$ and $K\overline{K}$ correlations ($R_{\Lambda\sigma}=0.49$)
\cite{Haidenbauer:1998kk}. The
scaling factor for the $\Lambda\Lambda\omega$ vertex 
at present is fitted heuristically to
hypernuclear data. The results for single $\Lambda$ hypernuclei are
very satisfactory, see fig.~\ref{fig:hnspec}.
The description is in quality --
on the percent level --  matching the result of purely phenomenological
mean-field models. The additional overhead in theoretical and mathematical
effort for our approach is,
however, paying off in a microscopic link to the free interaction.
An even more important theoretical aspect is the complete control on the dynamical
content represented by specific classes
of diagrams in the kernel of the DB equations.

Encouraged by the results, work on Dirac-Brueckner calculations
for the complete baryonic octet based on meson exchange potentials of the 
Bonn-J\"lich type is in progress. 
Together with high precision spectroscopic measurements on single-
and multi-hypernuclei, this offers the unique possibility to determine interactions
for the complete baryonic octet within a consistent microscopic frame.
For this approach -- based on the relativistic mean-field 
(DDRH theory) -- it will be,
however, necessary, that high precision measurements are pursued on higher mass
systems to obtain results as reliable as possible. The heavier 
systems are
much better suited for structure calculations because 
finite size effects occuring in light nuclei are minimized \cite{Keil:2001rm}.

\section{HBT correlations}
\begin{figure}
\centering\epsfig{file=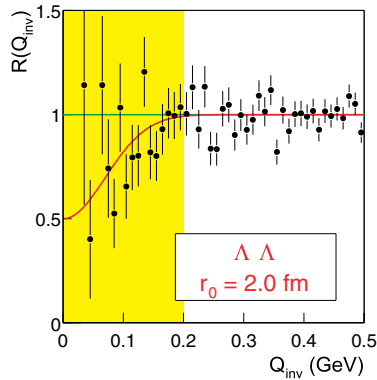,width=.4\linewidth}
\caption{HBT correlations in the $\Lambda\Lambda$ channel measured by NA49
\cite{Blume2001}. The
shaded area marks the relevant relative momenta influenced by strong interaction
between the $\Lambda$s}
\label{fig:HBT}
\end{figure}
A method complementing the one presented above for achieving information on
baryon-baryon interactions is the measurement of Hanbury-Brown-Twiss (HBT) two
particle correlations in very high energetic collisions of heavy ions. This was
studied first for hyperons ($\Lambda\Lambda$) in
\cite{Greiner:1989ig}. It was shown on
the example of a s-wave resonance in the $\Lambda\Lambda$ channel that
interactions between the two particles
modify quite significantly the standard correlations 
solely due to fermion quantum statistics,
namely resulting in a negative correlation of -0.5 when the relative 
momentum $q_{rel}$ approaches 0.
The HBT correlation function $R$ measures the ratio of the product of the two 
differential single particle cross sections to the differential two particle cross
section, each normalized to the respective total cross section:
\begin{equation}
\frac{1}{\sigma_{12}}\frac{d^6\sigma_{12}}{d\vec p_1d\vec p_2} =
\left( 1+R(q_{rel}) \right) 
\frac{1}{\sigma_{1}}\frac{d^3\sigma_{1}}{d\vec p_1}
\frac{1}{\sigma_{2}}\frac{d^3\sigma_{2}}{d\vec p_2}
\end{equation}
First measurements for $\Lambda\Lambda$ correlations have been performed by WA97
which, however, were of still rather poor statistics \cite{Andersen:1999gq}.
More encouraging preliminary results were obtained in a more recent analysis
by NA49 \cite{Blume2001}, see fig.~\ref{fig:HBT}. The error
bars, especially in the relevant
region below 200 MeV relative momentum, are still
too large due to insufficient statistics.

Within the same meson exchange model as described in the above section
these correlations can be computed by solving the
T-matrix equation
$
{\mathcal T}(q',q,s) = {\mathcal V}(q',q,s) + 
  \int\; d^4k{\mathcal V}(q',k,s) G^{(2)}(k,s){\mathcal T}(k,q,s)
$
and then compared to
data. In the case of sufficently high statistics 
this will give detailed information
on the interactions of the two particles. More accurate data is likely to be taken at
RHIC in the near future. However, there are still some ``impurities''
in the extraction of interaction informations in that way. First, the experimental
spectra include secondary $\Lambda$s from the 
electro-magnetic decay of initially produced
$\Sigma^0$ hyperons which thus have to be accounted for
in the calculated correlations. They
will somehow wash out the correlation structure of 
the pure $\Lambda\Lambda$ channel \cite{Greiner:1989ig}. 
Secondly, the source function of the $\Lambda$s (and the $\Sigma^0$s) 
needs to be modelled as this will influence the width of the 
momentum region in which
correlations are pronounced.

This work was supported by the ``European Graduate School: Complex Systems of
Hadrons and Nuclei, Gie\ss en--Copenhagen'' and GSI Darmstadt.


\end{document}